\documentclass[aps,prl,epsfigure,showpacs,twocolumn]{revtex4}
\usepackage{graphicx}
\usepackage{epsfig}
\usepackage{graphics}
\usepackage{amsmath}
\usepackage{amstext}
\usepackage{latexsym}
\usepackage{amsfonts}
\usepackage{amssymb}
\usepackage{array}
\usepackage{bm}

\newcommand{\ket}[1]{\left\vert#1\right\rangle}
\newcommand{\bra}[1]{\left\langle#1\right\vert}
\begin{document}

\title{Generation of a coherent superposition state on demand }
\author{M. S. Kim and M. Paternostro}
\affiliation{
  School of Mathematics and Physics, Queen's University, Belfast BT7
  1NN, United Kingdom
  }

\date{\today}

\begin{abstract}
We propose an experimentally feasible scheme to generate a superposition of travelling field coherent states using extremely small Kerr effect and an ancilla which could be a single photon or two entangled twin photons. The scheme contains ingredients which are all within the current state of the art and is robust against the main sources of errors which can be identified in our setups.
\end{abstract}

\pacs{03.67.Mn,42.50.Dv,42.50.Xa }

\maketitle

It has been a long dream for many physicists to produce a superposition of coherent states $|\pm\gamma\rangle$ which are $180^\circ$ out of phase with respect to each other
\begin{equation}
|\Psi\rangle_{\pm}={\cal N}_\pm\left(|\gamma\rangle\pm|-\gamma\rangle\right)
\label{cat}
\end{equation}
with normalization factors ${\cal N}_{\pm}$. The motivations of such the dream are manifold.  Firstly, among the pure states, this is arguably the most closely related to the Schr\"odinger's cat paradox~\cite{Schro}, which is one of the most profound and elegant examples of controversial behavior of the quantum world. In the paradox, a superposition of two macroscopic classical objects are quantum mechanically superposed. A coherent state is meant to describe a field which is classically considered to be stable light~\cite{Loudon} and a superposition of two distinct coherent states has been theoretically shown to manifest various nonclassical properties~\cite{buzek}. Secondly, the coherent superposition state is the most important ingredient for many applications of quantum information processing (QIP) with a coherent state, including teleportation~\cite{p}, computation~\cite{Jeong} and precision measurement~\cite{brau}.  

Despite some success in creating such a state  within high quality-factor cavities in the microwave~\cite{Brune96} and in trapped -ion systems in the optical domains~\cite{Monroe}, it has always been seen as an extremely challenging task to generate the free propagating coherent superposition state in Eq.~(\ref{cat}) while nearly all the coherent QIP applications are proposed for the propagating field. Recently, a conditional production of coherent superposition states was studied~\cite{Welsch,Lund}. Also, mechanisms based on the use of giant engineered nonlinearity via particular interaction regimes between matter and light ({\it i.e.} electromagnetically induced transparency) have been put forward in order to generate nonclassical states of travelling fields~\cite{eit}. 

As early as 1986, Yurke and Stoler, in their seminal work, proposed to generate a coherent superposition state using a Kerr nonlinearity~\cite{Yurke}, whose interaction Hamiltonian is $\hat{H}_0=\hbar\lambda_0\hat{n}_a^2$, where $\lambda_0$ is the rate of nonlinearity and $\hat{n}_a$ is the photon number operator of the field mode. However, this has never been realized because of the rate $\lambda_{0}$ being extremely low. While the nonlinear coupling $\lambda_0 t$ with its coupling time $t$ has to be about $\pi/2$, the currently achievable experimental value is less than one hundredth of it. 

In this paper, we propose an experimentally feasible scheme to produce a free propagating coherent superposition state with extremely low Kerr nonlinearity.  Another important point of our proposal is that its success is not conditional so that we get the coherent superposition state at every run.  This may be compared with other quantum manipulation schemes~\cite{KLM} where high nonlinearity is generated conditionally, requiring far too many resources so that the corresponding efficiency has been largely questioned.   

Inspired by the quantum nondemolition measurement~\cite{braginsky}, recently Nemoto and Munro~\cite{munro} have proposed a scheme to perform quantum computation using the cross Kerr nonlinearity, whose evolution operator is given by
$\hat{U}_{cK}=\mbox{e}^{i\phi\hat{n}_a\hat{n}_b}$,
which affects the phase of system depending on the photon numbers of two modes $a$ and $b$ ($\hat{n}_b$ is the photon number operator for mode $b$.). Here the parameter $\phi$ has been defined as the corresponding cross-Kerr nonlinear rate $\lambda$ times the interaction time $t$: $\phi=\lambda t$. The scheme in~\cite{munro} also uses a coherent state as a bus for gate operations between logical qubits of horizontal $|1_H\rangle$ and vertical $|1_V\rangle$ modes of a single photon state. Let us assume that mode $a$ is in a coherent state of its amplitude $\alpha$ and mode $b$ in a single photon state of vertical or horizontal polarization, then 
as the coherent state $|\alpha\rangle=\mbox{e}^{-|\alpha|^2/2}\sum_{n=0}^\infty (\alpha^n/\sqrt{n!})|n\rangle$, 
by the action of the cross-Kerr nonlinear coupling, the system of modes $a$ and $b$ becomes to be in 
\begin{equation}
\hat{U}_{cK}|\alpha\rangle_a|1\rangle_b=|\alpha\mbox{e}^{i\phi}\rangle_a|1\rangle_b,
\label{change}
\end{equation}
regardless of the polarization of $b$. By the action of the cross-Kerr coupling, the coherent state is rotated by $\phi$ in phase space.

\begin{figure} [ht]
\vskip-0.6cm
\psfig{figure=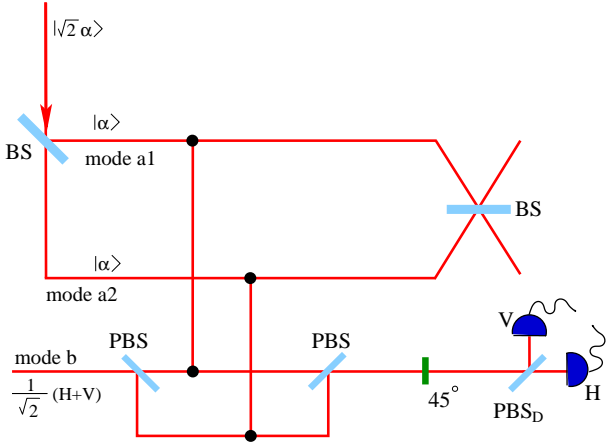,width=8.0cm,height=5.9cm}
\caption{Schematic diagram to generate a coherent superposition state using a single photon and a strong coherent field with an extremely low Kerr nonlinearity. The lines with two circles at the ends denote Kerr interaction. BS: beam splitter. PBS: polarization beam splitter. A $45^\circ$ polarization rotator and two photodetectors are also shown.}  
\label{fig1}
\end{figure}

Our proposal to generate a coherent superposition state is depicted in Fig.~1.    First,  for mode $b$ we prepare a single photon state in a superposition of horizontal and vertical polarizations as $({1}/{\sqrt{2}})(|1_H\rangle+|1_V\rangle)$. A coherent field of amplitude $\sqrt{2}\alpha$ is splitted equally into two modes $a1$ and $a2$ by a 50:50 beam splitter (BS).  Modes $a1$, $a2$ and $b$ are then initially in ${\frac{1}{\sqrt{2}}}|\alpha, \alpha\rangle_{a1,a2}(|1_H\rangle_b+|1_V\rangle_b)$.  If the single photon is horizontally polarized, it will pass through the polarization beam splitter (PBS) and cross-Kerr interact with mode $a1$.  This will rotate the phase of the coherent state in mode $a1$ as shown by Eq.~(\ref{change}).  Analogously, if the single photon is vertically polarized, it will be reflected by the PBS and mode $a2$ instead of $a1$ will be rotated.  The state of the all three modes is then
\begin{equation}
\label{perfetto}
{\frac{1}{\sqrt{2}}}\left(|\alpha^\prime,\alpha,1_H\rangle_{a1,a2,b}+|\alpha,\alpha^\prime,1_V\rangle_{a1,a2,b}\right),
\end{equation}
where $\alpha^\prime=\alpha\mbox{e}^{i\phi}$.  By the 45$^\circ$ rotation of the polarization for the single photon, $|1_H\rangle\rightarrow{\frac{1}{\sqrt{2}}}(|1_H\rangle+|1_V\rangle)$ and $|1_V\rangle\rightarrow{\frac{1}{\sqrt{2}}}(|1_H\rangle-|1_V\rangle)$.  Thus after the 45$^\circ$ rotation, the total modes are in
\begin{equation}
\label{single-final}
\begin{aligned}
&\frac{1}{2}\left[(|\alpha^\prime,\alpha\rangle_{a1,a2}+|\alpha,\alpha^\prime\rangle_{a1,a2})|1_H\rangle_b\right.\\
&\left.+(|\alpha^\prime,\alpha\rangle_{a1,a2}-|\alpha,\alpha^\prime\rangle_{a1,a2})|1_V\rangle_b\right].
\end{aligned}
\end{equation}
After PBS$_{\mbox{D}}$ we measure the polarization of the photon in mode $b$. If the photon is detected in horizontal polarization we know that modes $a1$ and $a2$ are in a so-called entangled coherent state 
\begin{equation}
|\Phi_H\rangle_{a1,a2}={\cal N}^\prime_+(|\alpha^\prime,\alpha\rangle_{a1,a2}+|\alpha,\alpha^\prime\rangle_{a1,a2}).
\label{hori}
\end{equation}
If the photon is vertically polarized, another entangled coherent state is produced
\begin{equation}
|\Phi_V\rangle_{a1,a2}={\cal N}^\prime_-(|\alpha^\prime,\alpha\rangle_{a1,a2}-|\alpha,\alpha^\prime\rangle_{a1,a2}),
\label{vert}
\end{equation}
where ${\cal N'}_\pm$ are normalization factors.  The class of entangled coherent states is {\it per se} very useful and interesting~\cite{Jeong,eit}. Here, however, we do not want to overemphasize the production of these states. As $\phi\ll 1$, unless $\alpha$ is very large, the overlap $|\langle\alpha|\alpha^\prime\rangle|$ is nearly unity so that the  characteristic properties of entanglement appears only for a large $\alpha$.  In this case, the decoherence is extremely fast as it depends on $\alpha$ \cite{buzek} and for such a large amplitude, revealing a quantum effect is not a trivial task~\cite{erlangen}.  

Now, let us consider the case of the single photon detected in horizontal polarization.  Taking modes $a1$ and $a2$ of $|\Phi_H\rangle_{a1,a2}$ to two input ports of a BS, we find at the two output ports $o1$ and $o2$ the state
\begin{equation}
{\cal N'}_+\left|\frac{\alpha+\alpha^\prime}{\sqrt{2}}\right\rangle_{o1}\left(\left|\frac{\alpha-\alpha^\prime}{\sqrt{2}}\right\rangle_{o2}\!+\left|\frac{-\alpha+\alpha^\prime}{\sqrt{2}}\right\rangle_{o2}\right),
\end{equation}
where $|\frac{\alpha-\alpha^\prime}{\sqrt{2}}\rangle=|{\frac{\alpha}{\sqrt{2}}}(1-\cos\phi)-i{\frac{\alpha}{\sqrt{2}}}\sin\phi\rangle$. We take a reasonable value for $\phi\approx2\times10^{-2}$ without having to consider dispersion~\cite{Jeong,munro} and, in this limit, we use the Taylor expansion of the trigonometric functions. By the approximation ${\cal O}(\phi^2)$, $|(\alpha-\alpha^\prime)/\sqrt{2}\rangle\approx|-i\alpha\phi/\sqrt 2\rangle\equiv|\gamma\phi\rangle$, where we assumed that $-i\alpha/\sqrt{2}=\gamma$ and $\alpha$ is imaginary ($\gamma$ real) without a loss of generality. Similarly, $|(\alpha^\prime-\alpha)/\sqrt{2}\rangle\approx|-\gamma\phi\rangle$. We have successfully generated a coherent superposition state $|\Psi_+\rangle$ in Eq.~(\ref{cat}) at the outport 2 of the BS (Fig.~\ref{fig1}).  Analogously, when the single photon is detected in vertical polarization, a coherent superposition state $|\Psi_-\rangle$ is generated.  

For the QIP application, we need $\gamma$ around 2 for which $|\alpha|\approx 10^{2}$ is required.  As a typical laser field has its intensity higher than $10^5$ per mode~\cite{Loudon}, this is easily achievable by putting a neutral density filter.  Any negativity in the Wigner function is taken as an important signal for the nonclassicality of the field~\cite{Loudon}. When $\gamma$ is 2, the Wigner function at the origin of the phase space is around -0.5~\cite{buzek}, which is already a very significant value.  Our condition may be compared with Nemoto and Munro's condition, $\gamma\phi^2\gg 1$ for the success of their qubit operations~\cite{munro}.

\begin{figure} [b]
\vskip-0.6cm
\psfig{figure=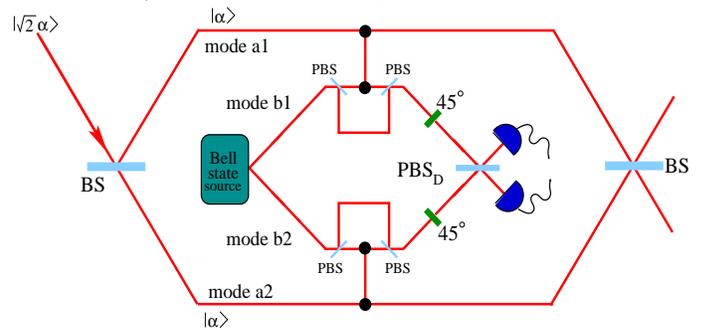,width=9cm,height=4.3cm}
\caption{Coherent superposition state generated by an entangled pair of photons and a low Kerr nonlinearity. The legend is the same as Fig.~\ref{fig1}.}  
\label{fig2}
\end{figure}

In Fig.~\ref{fig2}, we show another scheme to generate a coherent superposition state.  This scheme shares the same spirit as the one described in Fig.~\ref{fig1} but, instead of a single photon, an entangled twin photons, which is becoming a standard device at a quantum optics lab, are used.  Consider that the the twin photons are created in a singlet state $|\varphi_s\rangle=(|1_H, 1_V\rangle_{b1,b2}- |1_V, 1_H\rangle_{b1,b2})/\sqrt{2}$ of modes $b1$ and $b2$ and a coherent field is beam splitted equally into modes $a1$ and $a2$, then before their nonlinear interaction, the state for the total system is
$|\varphi\rangle_{b1,b2}|\alpha,\alpha\rangle_{a1,a2}$.
The singlet resource may be generated by using, for instance, a Type II spontaneous parametric down conversion process, which is nowadays routinely implemented in laboratory~\cite{kwiat}. By the action of the nonlinear couplings, the state becomes
$
(|1_H\rangle_{b1}|\alpha^\prime\rangle_{a1}|1_V\rangle_{b2}|\alpha\rangle_{a2}-|1_V\rangle_{b1}|\alpha\rangle_{a1}|1_H\rangle_{b2}|\alpha^\prime\rangle_{a2})/\sqrt{2}
$.  After the 45$^\circ$ rotation of the polarization for each of the twin photons as shown in Fig.~2, the state transforms to  
\begin{eqnarray}
&\frac{1}{\sqrt{2}}|\varphi_t\rangle_{b1,b2}\left[(|\alpha^\prime,\alpha\rangle_{a1,a2}
-|\alpha,\alpha^\prime\rangle_{a1,a2})\right.\nonumber \\
&~~~~\left.-|\varphi_s\rangle_{b1,b2}(|\alpha^\prime,\alpha\rangle_{a1,a2}\right]
+|\alpha,\alpha^\prime\rangle_{a1,a2})
\label{2-1}
\end{eqnarray}
where $|\varphi_t\rangle=(|1_H,1_H\rangle-|1_V1_V\rangle)/\sqrt{2}$. We then measure twin photons after they pass through a PBS.  By counting photons only in one of the detectors, we know that the field in modes $a1$ and $a2$ is in state $|\Phi_H\rangle$ of Eq.~(\ref{hori}) and by counting photons at the both photodetectors, it is in $|\Phi_V\rangle$. Now, the rest of the process is exactly the same as the case for Fig.~1.

In this paper, we have shown experimentally feasible schemes to generate an extremely nonclassical state.  In the single-photon implementation case, our scheme should be robust against the photon loss or the inefficiency of the photo detector as the success of the scheme is triggered by the photodetection.  If the single photon is lost during the process or at the detection, no photons will be detected and we have to start a new process.  Thus, this will not be a source of error even though it will cause inefficiency in the generation of the coherent superposition state. Obviously, in the twin-photon implementation case, as $|\Phi_H\rangle$ and $|\Phi_V\rangle$ are determined by not detecting a photon in one side, there is a chance that we wrongly predict the final state between $|\Psi_+\rangle$ and $|\Psi_-\rangle$.  For the case of dark counts, we have more serious problem but the contribution from the dark count can be reduced by properly designing the experiment, for example by narrowing the photodetection time window, choosing a right frequency and lowering the operating temperature. Another source of imperfection can come from the uncertain Kerr nonlinearity. With reference to Fig.~\ref{fig1}, for example, the two different Kerr interactions could last for not equal interaction times, so that the two modes $a1$ and $a1$ acquire different dynamical phases. Analogous reasoning holds for Fig.~\ref{fig2}, where the two different Kerr materials could induce this very same sort of imperfection. In order to fix the ideas, in what follows we refer to the scheme in Fig.~\ref{fig1}. This nonideality will lead us to the (un-normalized) entangled coherent state, after a detection which has revealed a horizontal polarization of the ancillary mode $b$,
\begin{equation}
\ket{\alpha',\alpha}_{a1,a2}+\ket{\alpha,\alpha''}_{a1,a2}.
\end{equation} 
Here, $\alpha'$ is defined as in Eq.~(\ref{perfetto}) while $\alpha''=\alpha{e}^{i\varphi}$ with $\varphi=\phi+\epsilon$ and $\epsilon$ is assumed to be a small difference between the two dynamical phases ({\it i.e.} $\epsilon\ll{\phi}$). After the second BS in Fig.~\ref{fig1} and regardless of mode $o1$, mode $o2$ is thus found to be in the state 
\begin{equation}
\label{density}
\begin{aligned}
\rho_{o2}&={\cal M}^{2}_{+}\left[\ket{\frac{\alpha-\alpha'}{\sqrt 2}}_{o2}\!\bra{\frac{\alpha-\alpha'}{\sqrt 2}}+\ket{\frac{\alpha''-\alpha}{\sqrt 2}}_{o2}\!\bra{\frac{\alpha''-\alpha}{\sqrt 2}}\right.\\
&\left.+\left(e^{i|\alpha|^2(\varphi-\phi)}\ket{\frac{\alpha-\alpha'}{\sqrt 2}}_{o2}\!\bra{\frac{\alpha''-\alpha}{\sqrt 2}}+h.c.\right)\right],
\end{aligned}
\end{equation} 
where ${\cal M}_{+}$ is a proper normalization function and $h.c.$ stands for hermitian conjugation. In order to test the robustness of our scheme to this imperfection, we have calculated the fidelity between $\rho_{o2}$ and the (properly normalized) desired superposition $|\frac{\alpha-\alpha^\prime}{\sqrt{2}}\rangle_{o2}+|\frac{-\alpha+\alpha^\prime}{\sqrt{2}}\rangle_{o2}$. For $\epsilon$ as large as $10\%$ of $\phi$, the fidelity is still larger than $0.95$, revealing a good resilience of the protocol to asymmetries in the Kerr interactions.

In conclusion, we have shown simple schemes for the deterministic generation of a coherent superposition of two macroscopically distinguishable states of traveling wave fields. Both the suggested schemes rely on experimentally achievable requirements and are, thus, immediately realizable in optical laboratories. Once a coherent superposition state is produced as an off-line resource coherent quantum information processing can be implemented with only linear optical components.

\acknowledgments 

We thank Dr. H. Jeong and Prof. J. Lee for discussions and for bringing our attention to Ref.~\cite{munro}. Financial supports from the UK EPSRC, the KRF (2003-070-C00024) and the Leverhulme Trust (ECF/40157) are acknowledged.

\end{document}